\documentclass[12pt]{article}
\usepackage{graphicx}
\usepackage{amsmath, amssymb}
\usepackage{hyperref}
\usepackage{authblk}
\usepackage{geometry}
\title{Machine Learning-Assisted Optimization of Modular Neutron Shielding Based on Monte Carlo Simulations}
\author[1, 2]{Li-Fang Chen\thanks{Corresponding author: lifangchen0507@gmail.com}}
\affil[1]{Department of Physics, National Taiwan Normal University, Taipei, Taiwan}
\affil[2]{National Atomic Research Institute (NARI), Taoyuan, Taiwan}
\date{}

\begin{document}
\maketitle

\begin{abstract}
This study proposes a novel design methodology for neutron beam shutters that integrates Monte Carlo simulations (MCNP) with machine learning techniques to enhance shielding performance and accelerate the design process. The target facility is a compact neutron science platform where neutrons are produced by proton beams from a cyclotron striking a neutron production target. A beam shutter is installed on the thermal neutron line to reduce occupational radiation exposure during maintenance activities.

In this work, 200 neutron shutter configurations with varying material sequences were simulated using MCNP. The resulting dataset was used to train a fully connected neural network to predict the neutron flux downstream of the shielding. The trained model was subsequently applied to 1,000 randomly generated shielding configurations for rapid flux prediction and performance ranking. The 20 designs with the lowest predicted flux were selected and further validated via MCNP simulations.

Results show that the optimal design reduces the neutron flux from $5.61 \times 10^9$ n/cm$^2{\cdot}$s at the shutter entrance to $4.96 \times 10^5$ n/cm$^2{\cdot}$s at the exit, achieving a reduction of four orders of magnitude. These findings confirm that the integration of machine learning techniques can effectively reduce simulation costs and assist in identifying high-performance shielding configurations, demonstrating the strong potential of data-driven approaches in neutron system design.


\medskip

\end{abstract}

\section{Introduction}
A compact neutron source system based on a cyclotron accelerator has been established in Taiwan, where high-energy protons impinge on a neutron-generating target to produce neutrons. Originally developed for radiopharmaceutical production, the facility has gradually expanded into a multifunctional platform for neutron-based research and applications.

The design and operation of such facilities require comprehensive simulations—ranging from proton beam transport and target configuration to neutron propagation and radiation shielding—to ensure performance and safety in compliance with international standards. A representative simulation study previously evaluated the cadmium ratio in a thermal neutron radiography system at this facility [1].

 The facility currently operates two major neutron beamlines: one for thermal neutrons and one for fast neutrons. The latter has recently incorporated a quasi-monoenergetic neutron (QMN) source driven by a 30~MeV cyclotron. Utilizing a beryllium target and supported by Monte Carlo simulations[2, 3], this system has demonstrated promising neutron yield and engineering feasibility through detailed thermal and flux modeling[4]. To ensure radiation safety during maintenance or beam-off conditions, a beam shutter is currently under planning and design for the thermal neutron beamline to block neutron transmission and reduce occupational exposure.

As a critical component of the neutron beam control system, the beam shutter must meet several engineering requirements, including high shielding efficiency, mechanical modularity, and structural replaceability. During shutdown states, the shutter must provide reliable neutron attenuation to safeguard downstream instruments and maintenance personnel. However, the configuration and layering of shielding materials exhibit highly nonlinear effects on both neutron flux and energy spectra. These challenges are exacerbated in mixed radiation fields, where both neutrons and photons are present. Given the significant differences in attenuation behavior and interaction mechanisms across particle types, shielding design must account for spectral characteristics of all radiation species. As such, relying solely on empirical approaches is insufficient for identifying optimal shielding configurations. Recent studies have investigated beam shutter design strategies in various applications [5–7], offering useful guidance for future development.

To date, most shutter designs rely on Monte Carlo methods, such as MCNP, for evaluating shielding performance and conducting parameter sweeps. For instance,  Didi, et al. designed a multilayered beam shutter for the Prompt Gamma Neutron Activation Analysis (PGAA) system at the TRIGA Mark II reactor and employed MCNP simulations to model the neutron attenuation and energy spectral changes through the shutter, highlighting its importance for radiation safety during system maintenance [8]. Although such simulations are highly accurate, their computational cost and time requirements become substantial when exploring high-dimensional design spaces, thereby limiting optimization efficiency.

In recent years, machine learning (ML) has emerged as a powerful tool in scientific computing, particularly in applications requiring the extraction of complex patterns from large datasets. In their landmark review, LeCun et al. introduced deep learning as a multi-layered nonlinear transformation framework capable of learning abstract feature representations from high-dimensional data, thus providing a theoretical foundation for its application in engineering and physical modeling [9]. Alzubaidi et al. summarized the evolution of convolutional neural network (CNN) architectures in image recognition and emphasized their strong generalization capability in predictive modeling tasks [10]. Uddin et al. provided a taxonomy of supervised, unsupervised, and hybrid learning models, and reviewed their applications in medicine, materials, and cybersecurity [11]. Leveraging ML models enables the prediction of shielding performance from limited simulation data, facilitating rapid screening and accelerated optimization of neutron shielding designs.

This study proposes a machine learning-assisted framework for neutron beam shutter design, integrating MCNP simulation data with a fully connected neural network (FCNN). We first generated a training dataset comprising 200 randomly assembled multilayer shielding configurations. The trained model was then applied to predict neutron flux performance for 1,000 candidate designs, from which the best-performing cases were selected and re-validated through MCNP simulation. The results demonstrate that data-driven ML models can significantly reduce simulation cost and human effort, thereby offering high potential for efficient and intelligent shielding design.
\section{Methods}\label{}
\subsection{Design Parameters of the Neutron Beam Shutter}
The neutron beam shutter is positioned 170 cm downstream from the neutron target. The shielding layers are composed of polyethylene, borated polyethylene, iron, and lead, designed to attenuate neutron flux and radiation dose effectively. Following the concept of using standardized modular components, each shielding block was fabricated with dimensions of 7 cm in length, 13 cm in height, and 2 cm in width. A total of 25 blocks were assembled sequentially, with the final layer fixed as a 1 cm thick iron plate.

This final iron layer not only contributes to neutron attenuation but also serves as an activation target for neutron spectrum and flux characterization through neutron activation analysis (NAA). The overall shielding section of the shutter measures 49 cm in total length and is housed within a removable stainless steel frame. This modular frame allows for rapid disassembly and relocation to a high-radiation laboratory when necessary, enabling testing and replacement of shielding materials. This design supports both spectral shaping of the neutron beam and replacement of radiation-degraded shielding components.

\begin{figure}[htbp]
  \centering
  \includegraphics[width=0.5\textwidth]{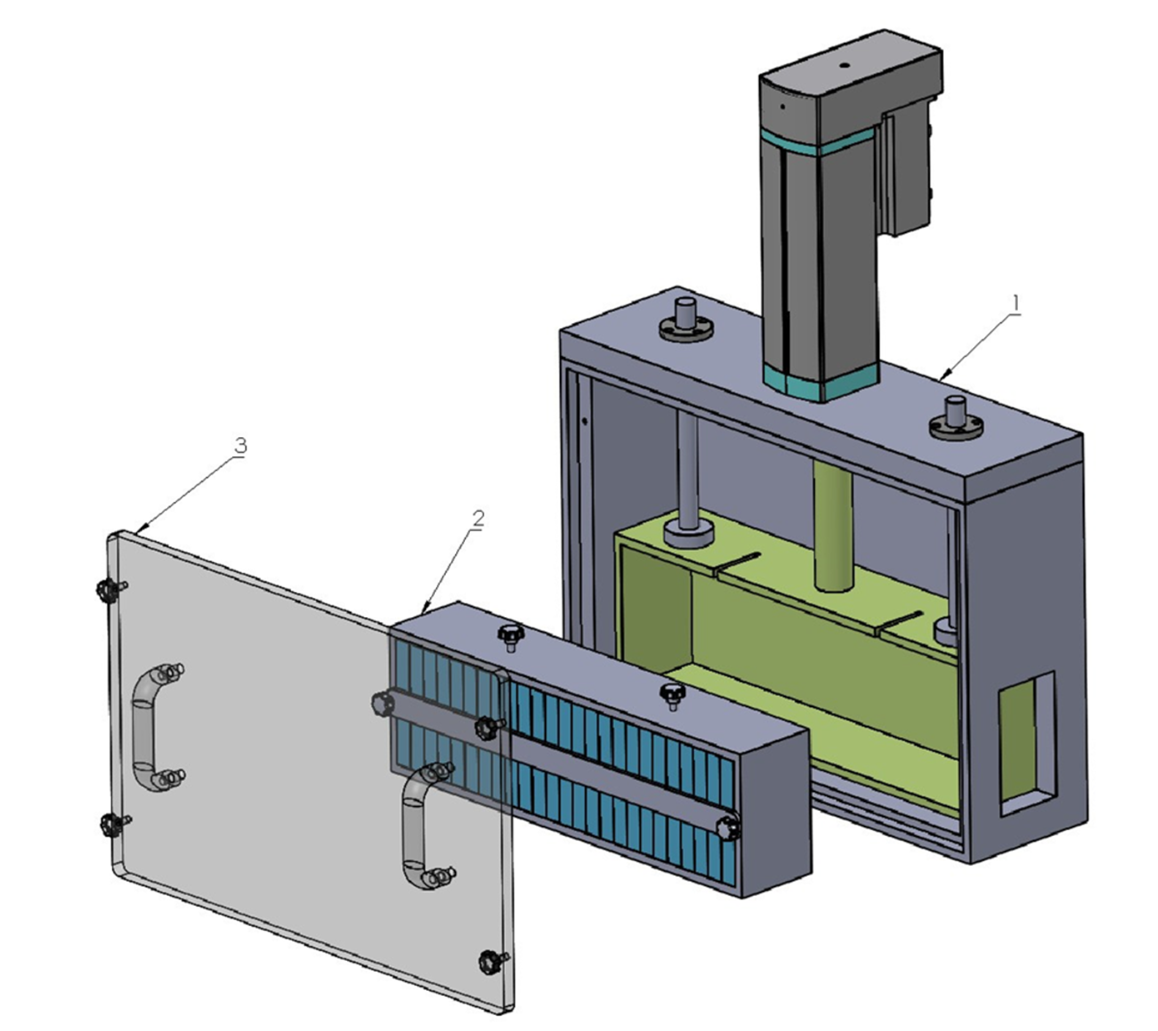}
  \caption{Schematic diagram of the modular beam shutter design. The figure shows the mechanical layout of the beam shutter system, including the actuator assembly (1), modular shielding block array (2), and detachable rear interface plate (3). The design supports modular maintenance and material replacement under high-radiation operation conditions.}
  \label{fig:beamshutter}
\end{figure}

Figure~\ref{fig:beamshutter} presents a simplified schematic of the beam shutter assembly. The design includes a modular shielding cartridge mounted inside a removable steel frame, with an access panel for maintenance and replacement. The vertical actuator mechanism (component 1) controls shutter movement, while the shielding cartridge (component 2) holds the 25-layer arrangement. The transparent plate (component 3) illustrates the rear access interface for shutter replacement and testing.

\subsection{MCNP Simulation Design and Data Generation}
In this study, neutron flux simulations were carried out using MCNP 6.2. The neutron source was reconstructed based on the flux and energy spectrum measured at the entrance of the neutron guide. The simulation geometry incorporated a vacuum neutron guide constructed from 5 mm-thick 6061 aluminum, with a cross-section of 6 cm × 12 cm, and a total length of 170 cm. The beam shutter was positioned at the end of the guide and consisted of modular shielding blocks encased within a 5 mm stainless steel frame, as described in Section 2.1. The geometric configuration is illustrated in Figure 1.
    
The MCNP output parameters included neutron flux and energy spectra at various positions along the beamline. The key performance indicator of the beam shutter was the neutron flux within a 1 cm-thick air layer located directly behind the shutter. A lower flux value in this region was interpreted as higher shielding effectiveness.

For flux and spectral calculations, the F4 tally was employed to record neutron flux and perform energy binning. The energy range from$10^{-8}$ MeV MeV to 30 MeV was divided into 15 energy groups: for energies below 1 MeV, each group corresponds to a single order of magnitude, while for energies above 1 MeV, the groups were divided in 5 MeV intervals. In addition to neutron flux, absorbed dose in air behind the shutter was also calculated using the F6 tally, providing further insight into shielding effectiveness from a radiation protection perspective.

\subsection{Machine Learning Approach for Beam Shutter Prediction}
To predict the shielding performance of various neutron beam shutter configurations, this study employed a supervised learning approach based on a fully connected deep neural network (DNN). The input features were defined as a 25-dimensional vector representing the material sequence of the 25 shielding layers within the beam shutter. Each shielding material--polyethylene, borated polyethylene, iron, and lead--was numerically encoded and mapped into the input space. The output target was defined as the total neutron flux (unit: cm$^{-2}\cdot$s$^{-1}$) in the air layer immediately downstream of the shutter, as determined by MCNP simulations.

A total of 200 unique beam shutter configurations were randomly generated and evaluated using MCNP 6.2 to compute the corresponding neutron flux values. Among these, 170 samples were used for training and validation, while the remaining 30 samples were reserved for testing. No cross-validation or early stopping strategies were applied in this implementation.

The neural network model was developed using the Keras API within the TensorFlow framework. The architecture consisted of three hidden layers with 128, 64, and 32 neurons, respectively. Each hidden layer utilized the ReLU activation function and was followed by batch normalization and dropout layers (with a dropout rate of 0.3) to mitigate overfitting. The output layer comprised a single neuron with no activation function, appropriate for regression tasks. The model was optimized using the Adam algorithm with a learning rate of 0.001, minimizing the mean squared error (MSE) as the primary loss function. Mean absolute error (MAE) was used as a secondary performance metric.

\section{Results}\label{}
\subsection{Simulation Data Analysis}
\begin{figure}[htbp]
    \centering
    \includegraphics[width=0.5\textwidth]{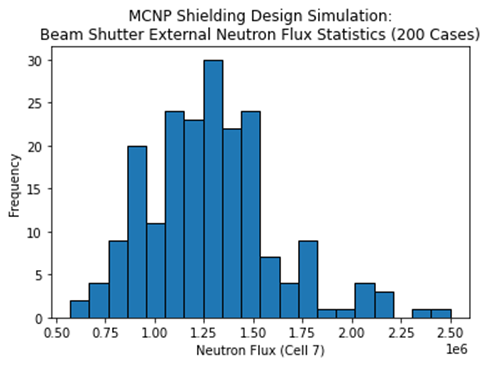}
    \caption{Distribution of total neutron flux behind the beam shutter across 200 MCNP simulations. The flux values exhibit a Poisson-like pattern, with a mode around $1.3 \times 10^{6}$~n/cm$^{2} \cdot$~s and a range from $5.7 \times 10^{5}$ to $2.5 \times 10^{6}$~n/cm$^{2} \cdot$~s.}
    \label{fig:flux_distribution}
\end{figure}

Statistical analysis was conducted on the total neutron flux obtained from 200 MCNP-simulated shielding configurations. As shown in Figure~\ref{fig:flux_distribution}, the flux values behind the beam shutter follow an approximately Poisson-like distribution, with the mode centered around $1.3 \times 10^{6}$~n/cm$^{2}\cdot$s, and a full range spanning from $5.7 \times 10^{5}$ to $2.5 \times 10^{6}$~n/cm$^{2}\cdot$s. This distribution reflects the stochastic nature of neutron shielding performance across varied material arrangements.

\begin{figure}[htbp]
    \centering
    \includegraphics[width=0.5\textwidth]{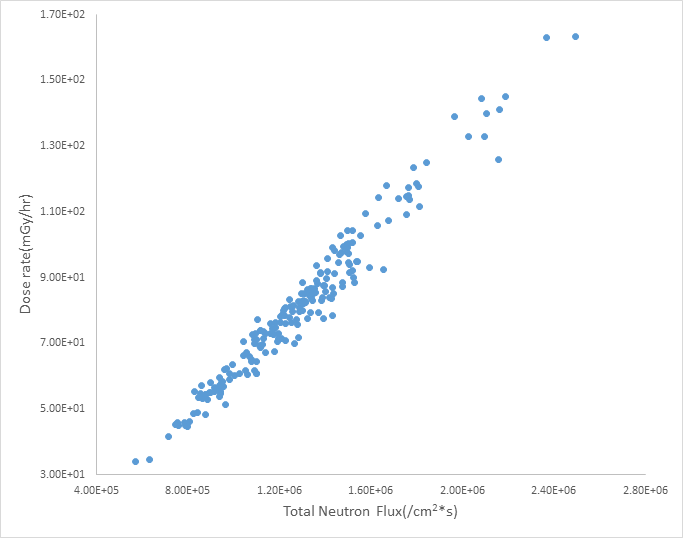}
    \caption{Distribution of total neutron flux behind the beam shutter across 200 MCNP simulations. The flux values exhibit a Poisson-like pattern, with a mode around $1.3 \times 10^{6}$~n/cm$^{2} \cdot$~s and a range from $5.7 \times 10^{5}$ to $2.5 \times 10^{6}$~n/cm$^{2} \cdot$~s.}
    \label{fig:Dose_vs_total}
\end{figure}

Figure~\ref{fig:Dose_vs_total} reveals a clear positive correlation between the total neutron flux and the neutron-induced absorbed dose in the air layer behind the shutter. While the total absorbed dose includes contributions from both neutrons and photons, only the neutron dose component is considered here due to its dominant magnitude.

In contrast, the relationship between thermal neutron flux ($E < 10^{-8}$~MeV) and total neutron flux, as presented in Figure~\ref{fig:Dose_vs_total}, shows no consistent trend across the 200 shielding configurations. This decoupling suggests that energy spectrum shaping may be achievable independently of total flux reduction by appropriately designing the shielding material sequence—offering additional flexibility in shutter optimization.

\begin{figure}[htbp]
  \centering
  \includegraphics[width=0.5\textwidth]{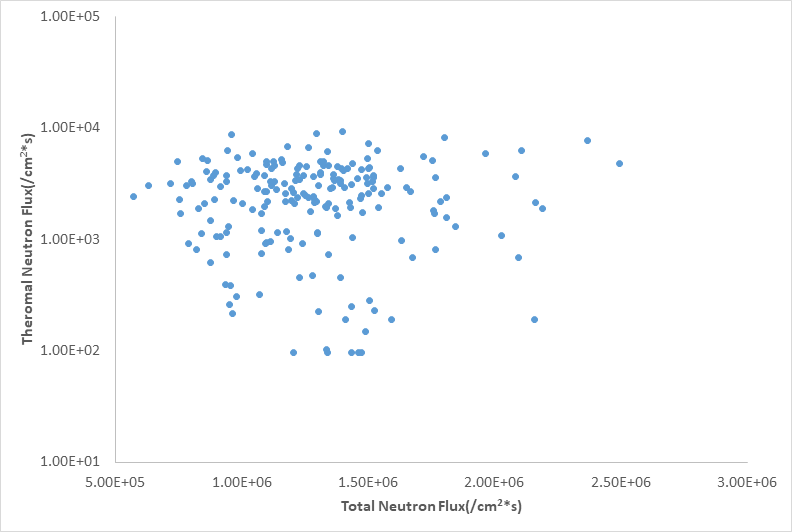}
  \caption{Scatter plot of thermal vs. total neutron flux across 200 shielding configurations. The lack of clear correlation suggests the potential to independently tailor the neutron spectrum via material design.}
  \label{fig:thermal_vs_total}
\end{figure}

\subsection{Machine Learning Model Performance}

A total of 200 MCNP simulation results were used to train and evaluate the performance of the neural network model. Among them, 170 samples were allocated to the training and validation sets, of which 80\% (136 samples) were used for training and 20\% (34 samples) for validation. The remaining 30 samples were reserved as an independent test set.

\begin{figure}[htbp]
  \centering
  \includegraphics[width=0.5\textwidth]{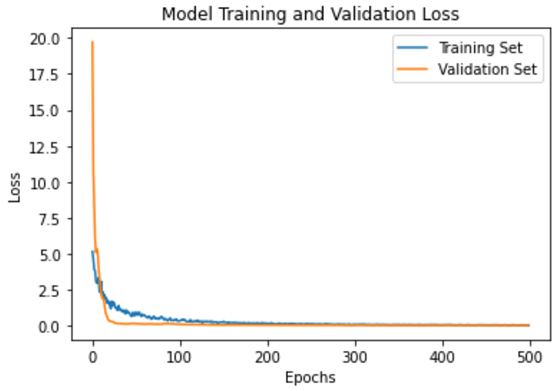}
  \caption{Training and validation loss curves over 500 epochs. The plot illustrates the training and validation loss (mean squared error) of the neural network model across 500 epochs. Both curves demonstrate a consistent decline and convergence trend, indicating stable learning and good generalization performance, with no signs of overfitting.}
  \label{fig:training_loss}
\end{figure}

The model was trained for 500 epochs using the mean squared error (MSE) as the loss function and the Adam optimizer. Figure~\ref{fig:training_loss} illustrates the evolution of training and validation losses throughout the learning process. Upon completion, the final training loss reached 0.0394 with a corresponding mean absolute error (MAE) of 0.1494, while the validation loss was 0.0211 with an MAE of 0.1097. These results indicate a stable training process with no apparent signs of overfitting.

Overall, the model exhibited consistent convergence in both squared and absolute error metrics, effectively capturing the nonlinear relationship between shielding configurations and neutron flux, and demonstrating good generalization performance.

\begin{figure}[htbp]
  \centering
  \includegraphics[width=0.5\textwidth]{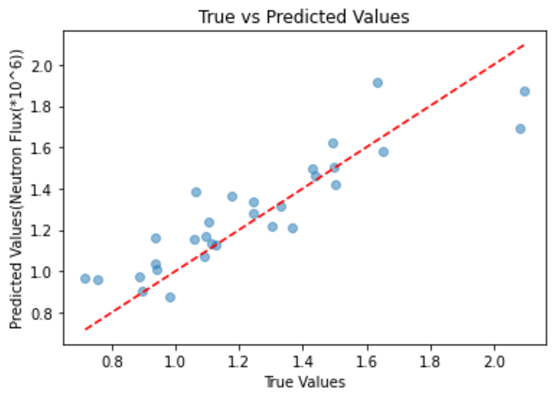}
  \caption{Comparison between neural network predictions and MCNP simulated neutron flux (30 test samples). The scatter plot shows predicted versus actual neutron flux values for the test set. The data points closely follow the ideal diagonal line, indicating strong agreement and high prediction accuracy.}
  \label{fig:prediction_vs_simulation}
\end{figure}

To evaluate the generalization performance of the model, predictions were performed on an independent test set consisting of 30 samples. Figure~\ref{fig:prediction_vs_simulation} shows the comparison between the predicted neutron flux and the corresponding MCNP simulation results. The predicted values exhibit a strong correlation with the ground truth, indicating the model's capability to accurately capture the underlying shielding behavior. The root mean squared error (RMSE) for the test set was calculated to be 0.0241, further confirming the effectiveness of the trained neural network.

\subsection{Beam Shutter Design Optimization}
To demonstrate the practical utility of the trained neural network, the model was applied to a newly generated dataset comprising 1,000 randomly constructed beam shutter configurations. The trained model predicted the total neutron flux for each configuration, and the top 20 designs with the lowest predicted flux were selected as promising candidates for optimal shielding performance.

\begin{figure}[htbp]
  \centering
  \includegraphics[width=0.5\textwidth]{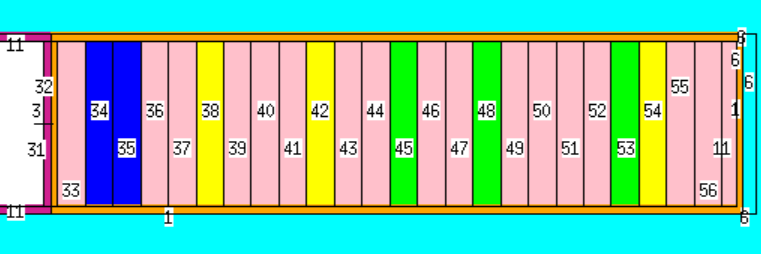}
  \caption{Geometric model of the optimal beam shutter design identified via neural network prediction and MCNP simulation. This figure presents the shielding configuration that achieved the lowest neutron flux among the top 20 machine learning–screened designs after MCNP re-evaluation. Different colors represent shielding materials: iron (pink), polyethylene (blue), borated polyethylene (yellow), lead (green), with a stainless steel frame (orange) and 6061 aluminum guide tube (magenta). The outer region is air (light blue). A total of 25 modular shielding layers are arranged as shown to achieve optimal neutron attenuation performance.}
  \label{fig:optimal_shutter_design}
\end{figure}

These 20 selected configurations were subsequently converted into MCNP input files and re-evaluated through full Monte Carlo simulations. Figure~\ref{fig:optimal_shutter_design} presents the shielding configuration that yielded the lowest neutron flux among the 20 re-simulated candidates. This result highlights the model’s effectiveness in pre-screening high-performance designs and significantly reducing the computational burden of optimization by restricting MCNP simulations to only the most promising configurations.

\begin{figure}[htbp]
  \centering
  \includegraphics[width=0.5\textwidth]{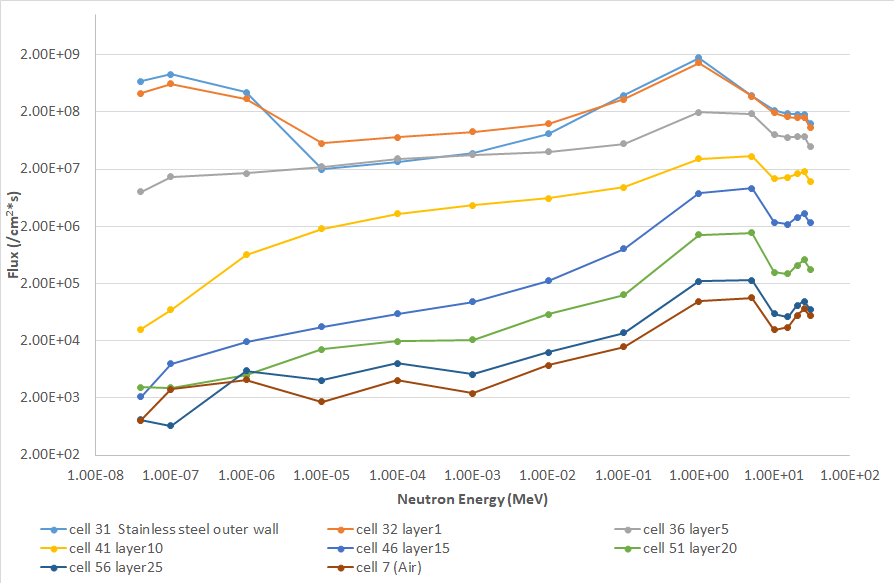}
  \caption{Neutron energy spectra and flux distribution at various depths within the optimized beam shutter. The figure illustrates the neutron flux spectra at multiple positions through the shielding layers (from cell 31 to cell 7). The $x$-axis represents neutron energy (MeV), and the $y$-axis shows flux in units of n/cm$^{2}\cdot$s. A clear downward trend in flux is observed across all energy ranges, with a notable reduction in thermal neutrons at layer 10, corresponding to a borated polyethylene layer. This confirms the design’s ability to reshape the neutron spectrum and effectively attenuate flux across multiple energy regions.}
  \label{fig:spectral_evolution}
\end{figure}

Figure~\ref{fig:spectral_evolution} illustrates the neutron energy spectra and corresponding flux distributions at various positions along the optimized beam shutter. The spectral evolution clearly reflects the impact of different shielding materials. A pronounced reduction in thermal neutron flux is observed at layer 10, corresponding to a borated polyethylene segment. This significant attenuation is attributed to the high thermal neutron absorption cross-section of boron, further enhanced by preceding layers that gradually moderate fast neutrons.

The overall shielding performance is remarkable: the total neutron flux decreases from $5.61 \times 10^{9}$~n/cm$^{2}\cdot$s at the shutter entrance to $4.96 \times 10^{5}$~n/cm$^{2}\cdot$s at the exit. This reduction spans nearly four orders of magnitude, clearly demonstrating the effectiveness of machine learning–assisted optimization in neutron beam shutter design.

\section{Discussion}\label{}

\subsection{Layer-wise Material Composition and Performance Interpretation}

\begin{figure}[htbp]
    \centering
    \includegraphics[width=0.95\textwidth]{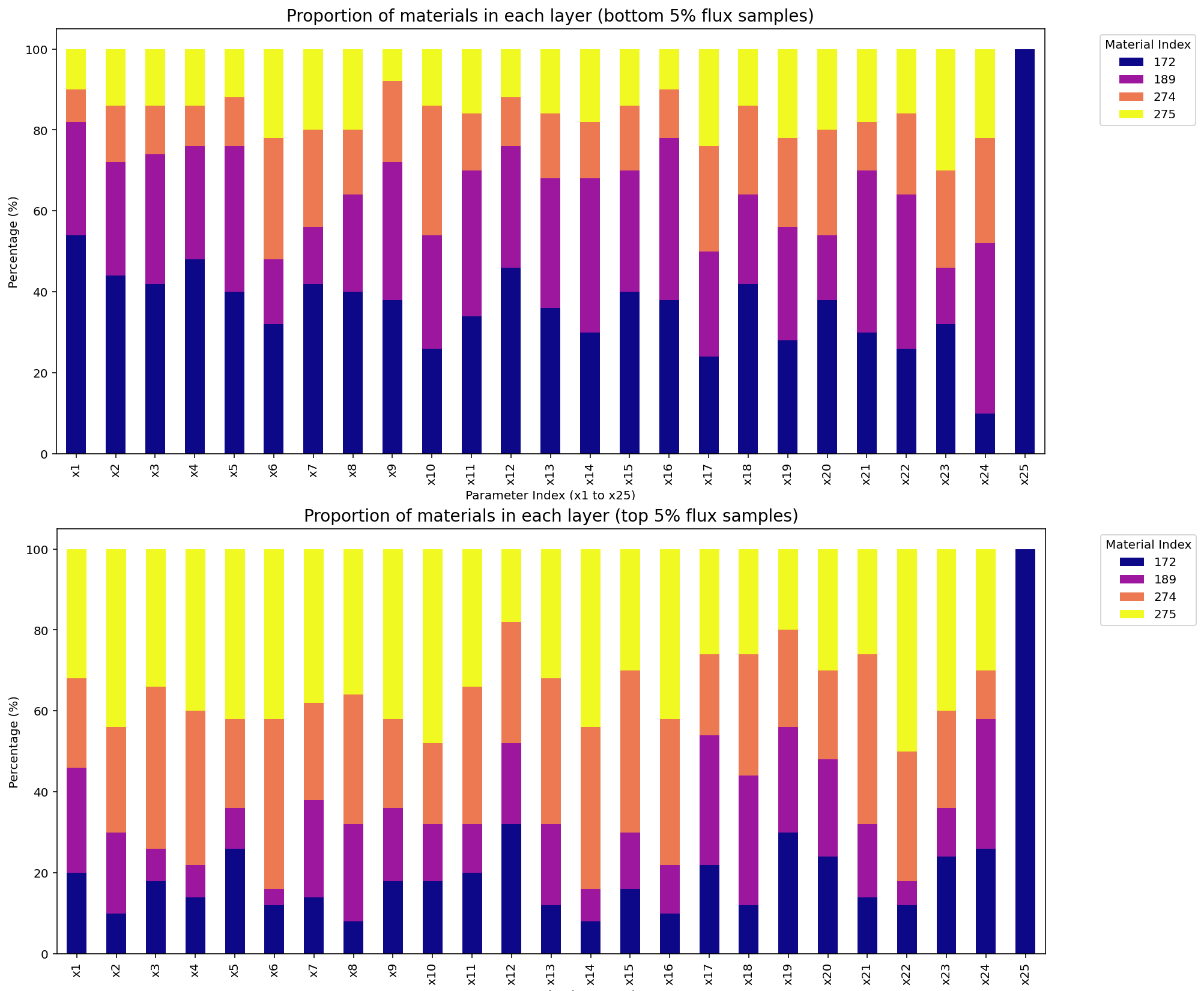}
    \caption{
    Layer-wise material composition for the top 5\% and bottom 5\% of shielding configurations, ranked by predicted total neutron flux from 1,000 randomly generated designs. The stacked bar charts indicate the percentage of each material at every layer. High-performance configurations exhibit a distinct preference for iron in the early layers, while low-performance cases show a uniform dominance of polyethylene-based materials. Material indices follow the classification in~\cite{ref:dmamc2021}.
    }
    \label{fig:layer_composition}
\end{figure}

Figure~\ref{fig:layer_composition} presents the statistical analysis of material composition across all 25 layers for the top and bottom 5\% of shielding configurations, ranked by total neutron flux predicted from 1,000 randomly generated designs using the trained machine learning model. Material indices shown in the figure correspond to the standardized classification scheme adopted from Reference~\cite{ref:dmamc2021}.

The analysis reveals that high-performance shielding configurations---those with the lowest predicted neutron flux---exhibit a strong preference for iron in the front layers (i.e., those closest to the neutron source), effectively moderating fast neutrons. Polyethylene and borated polyethylene are periodically distributed in the middle and rear layers, providing neutron moderation and absorption. In contrast, low-performance configurations (with the highest flux values) tend to contain high proportions of polyethylene and borated polyethylene across all layers, lacking sufficient high-density materials in the initial layers to attenuate fast neutrons.

These results suggest that although polyethylene-based materials offer favorable moderation and absorption properties, their low density limits their standalone effectiveness in comprehensive shielding. Strategic layering---introducing high-density materials in the early stages, followed by appropriately arranged moderator and absorber layers---is essential for achieving substantial reductions in neutron flux and enhancing overall shielding performance.

This study demonstrated a novel approach that integrates machine learning with MCNP simulation for the design and optimization of a neutron beam shutter. The neural network model, trained on 200 simulated configurations, achieved a strong predictive performance with a test set RMSE of 0.14, confirming its ability to accurately estimate neutron flux attenuation behind the shutter.

Further application of the trained model to a larger set of 1,000 randomly generated shielding designs enabled efficient pre-screening of candidate configurations. The top 20 designs with the lowest predicted flux were validated through MCNP re-simulation, and the results confirmed that the machine learning–assisted process can reliably identify high-performance shielding arrangements while significantly reducing computational cost.

In addition, the neutron spectrum evolution illustrated in Figure 8 reveals the shutter's capability in spectral shaping. A sharp decrease in thermal neutrons was observed at the borated polyethylene layer, supported by the prior moderation effects from earlier shielding materials. This layered design effectively reduces both fast and thermal neutron flux, achieving nearly four orders of magnitude attenuation.

Nonetheless, some limitations remain. The training dataset did not exhaustively cover all possible permutations of material sequences, and no cross-validation or hyperparameter optimization was applied. Future work may explore the use of Bayesian optimization, genetic algorithms, or reinforcement learning techniques to enhance model generalization and improve the global search for optimal shielding configurations.

\section{Conclusion}\label{}
This study successfully integrated Monte Carlo simulation and deep learning to develop a practical and scalable methodology for the design and optimization of neutron beam shutters. The concept of standardized modular shielding components was introduced, where each shielding layer was designed with unified dimensions and specifications. This approach provides the beam shutter with enhanced structural flexibility and ease of maintenance. It also facilitates regular inspection and replacement of radiation-sensitive or degraded materials, thereby extending system lifetime. Additionally, the modular design allows dynamic adjustment of the shielding configuration to shape the neutron spectrum as needed, while reducing long-term maintenance and fabrication costs.

By training a neural network model on 200 MCNP simulation cases, the system achieved accurate prediction of neutron flux attenuation across different shielding configurations, reaching a root mean squared error (RMSE) of only 0.14 on the test set.

When applied to a set of 1,000 randomly generated shielding designs, the trained model effectively pre-screened high-performance candidates. The best-performing configuration, verified by MCNP simulation, achieved a neutron flux attenuation of nearly four orders of magnitude.
The proposed framework demonstrates significant potential for efficient, cost-effective, and extensible shielding design. Future research may incorporate advanced techniques such as Bayesian optimization, cross-validation, and genetic algorithms to further improve model generalization and extend this methodology to other neutron instrument components, such as collimators, neutron guides, and spectrum filters.

\clearpage 

\section*{Declaration of generative AI and AI-assisted technologies in the writing process}

During the preparation of this work, the authors used ChatGPT (OpenAI) to improve the clarity and readability of the English language. After using this tool, the authors carefully reviewed and edited the content to ensure accuracy and take full responsibility for the final version of the manuscript.

\bibliographystyle{cas-model2-names}

\end{document}